\documentclass[%
 preprint, 
superscriptaddress,
 twocolumn,
 amsmath,amssymb,
 aps, physrev,
]{revtex4-2}

\usepackage{graphicx}
\usepackage{dcolumn}
\usepackage{bm}
\usepackage{hyperref}
\usepackage{orcidlink}  

\begin{document}

\preprint{DOI: \href{https://doi.org/10.1134/S1062873825711602}{10.1134/S1062873825711602}}

\title{\textbf{Modulation of the Galactic Cosmic Ray Spectrum in an Anisotropic Diffusion Approach} 
}%

\author{V.D. Borisov\,\orcidlink{0009-0005-2862-7133}}
\email[]{borisov.vd.19@physics.msu.ru}
\affiliation{Faculty of Physics, M.V. Lomonosov Moscow State University (MSU),
119991 Moscow, Russia}

\author{V.O. Yurovsky\,\orcidlink{0009-0008-1031-4499}}
\affiliation{Faculty of Physics, M.V. Lomonosov Moscow State University (MSU),
119991 Moscow, Russia}


\author{I.A. Kudryashov\,\orcidlink{0009-0009-1889-6232}}
\email[]{ilya.kudryashov.85@gmail.com}
\affiliation{Skobeltsyn Institute of Nuclear Physics, Lomonosov Moscow State University (MSU),
119991 Moscow, Russia}




\begin{abstract}
We introduce a novel diffusion model for the propagation of cosmic rays (CRs) that incorporates an anisotropic diffusion tensor of a general form within a realistically modeled large-scale Galactic magnetic field. The parameters of the model are consistent with the contemporary understanding of the large-scale Galactic magnetic field structure and the dynamics of small-scale turbulent CR propagation. The paper demonstrates the modulation of spectra of Galactic cosmic rays (GCRs) in the magnetic rigidity range of 1–30 PV (the CR knee) and explores the spatial variation of this phenomenon. The observed modulation of the spectrum is explained by changes in the leakage mechanism.
\end{abstract}

\maketitle


\section{\label{sec:level1}Introduction}

The experimentally observed spectrum of Cosmic Rays (CRs) exhibits distinct features across various energy ranges, including the cutoff at 3 PeV (for the proton component). Traditionally, features in the PeV energy range are described by introducing a break in the CR source spectrum, signifying an acceleration limit. The parameters of this break, such as the spectral slopes $\gamma_1$ before and $\gamma_2$ after the break, the smoothing factor, and the cutoff energy, are defined through the fit of the observed data.\par
Recent numerical experiments [1, 2] indicate significant anisotropy of the diffusion tensor and a strong dependency of its components on the magnitude of the large-scale Galactic magnetic field and the turbulence model considered. Therefore, it is crucial to account for significant modulation of the CR spectrum during their propagation from the source to the observer, along with other collective effects. Recent comprehensive studies including full-sky Faraday rotation measures of extragalactic sources and polarized synchrotron intensity (a detailed interpretation of the data is presented in [3]), along with investigations into changes in the CR leakage mechanism from the Galaxy [4], support the use of an anisotropic diffusion model to describe CR propagation.\par
The description of the free parameters of the model, including the structure of the large-scale magnetic field, the turbulence model, and the CR source distribution function, is provided. We described the numerical solution of the anisotropic diffusion equation in the developed model. The specifics of the observed spatial distribution of CR, as well as the nature of their outflow into the Galaxy’s halo, are discussed. Also, we discussed the modulation of the energy spectrum of CR protons in the energy range of 1–30 PeV and consider the spatial variation of the observed break.

\section{\label{sec:MODEL} MODEL}

In the software code developed, adjustments can be made to the model’s free parameters to explore various aspects of CR propagation. The model incorporates a classical description of the Galaxy’s large-scale magnetic field as detailed in JF12 [5] and further expanded in Unger2024 [3]. It considers the spiral structure of the arms, the variation of the magnetic field’s magnitude along each arm, as well as the diminishing magnetic field when moving in vertical and radial directions. We characterize the spiral structure of the Galaxy’s arms using the equations provided in [3]. \par

The diffusion tensor $\hat{D}(\mathbf{r},E,B)$ is computed at
each point, depending on the spatial position $\mathbf{r}$, energy of the particle $E$, magnitude of the magnetic
field vector $B (B_x,B_y,B_z)$ , with $B_x$, $B_y$ , and $B_z$ representing the projections of vector $B$ onto the Cartesian coordinate system’s axes. The plane $z = 0$ aligns with the Galaxy’s disk plane, and the coordinate system’s origin is at the Galaxy’s center. The values of the $\hat{D}$ tensor components vary for each instance of the turbulent field, reflecting the variation in energy dependency across different turbulent fields, as detailed in reference [1, 6]. Source distribution is defined according to the study referenced in [7]:

\begin{eqnarray}\label{eq:1}
S(R,z) =\left( \frac{R}{R_{sun}}\right)^a&&\times\nonumber\\ \exp\left(-b\frac{R - R_{sun}}{R_{sun}}-\frac{|z|}{z_0}\right)
\end{eqnarray}
where a = 1.9, b = 5, $R_{sun}$ = 8.2 kpc, and $z_{sun} = 0.2$ kpc is the is the position of the Sun. We assume that the sources are distributed on a thin disk with a thickness
of 0.3 kpc and a major semi-axis of 17 kpc.

\section{\label{MODEL_TRANSPORT_EQUATION} MODEL. TRANSPORT EQUATION}

The software package developed solves a second-order accuracy stationary system of partial differential
equations with a general form of anisotropic diffusion
tensor:

\begin{equation}\label{eq:1}
\begin{cases}
    \frac{\partial}{\partial x_i} \left( D_{ij}(\mathbf{r}) \frac{\partial f(\mathbf{r})}{\partial x_j} \right) = S(\mathbf{r}), \\
    f(\mathbf{r})|_S = 0,
\end{cases}
\end{equation}

where $S(\mathbf{r})$ represents the number density of sources at the point $\mathbf{r} ( x_1, x_2, x_3 )$, determined by Eq. (1), also here, summation over repeated indices is implied. Spatial boundary conditions allow for the free escape of particles from the Galaxy. \par

To determine the coefficients for the transport Eq. (2), a fully implicit finite difference method was selected to ensure the stability of the scheme. The spatial step, $x$, used to solve Eq. (2) was chosen to be larger than the maximum particle displacement necessary to achieve the diffusion mode of propagation at the specified energy. Estimates of $x$ were conducted in Study [1] for various field configurations. The minimum permissible step in calculations in the energy range under consideration is 100 pc.

\section{\label{RESULTS_CALCULATION_OF_CRs_NUMBER_DENSITY} RESULTS. CALCULATION
OF CR’s NUMBER DENSITY}

In this study, we have performed calculations of
cosmic ray proton (CRP) density across an energy
spectrum ranging from 1 TeV to 1 EeV. We employed
an implicit second-order accurate difference scheme
on a non-uniform grid, with a spatial resolution of
100 pc along the applicate axis and 0.56 kpc along
both the abscissa and ordinate axes of a Cartesian
coordinate system centered on the galaxy. At each energy, the diffusion tensor components $\hat{D}(\mathbf{r},E,B)$
were computed at grid nodes, utilizing the known values of the magnetic field’s magnitude and direction. The model for $\hat{D}(\mathbf{r},E,B)$ at each energy $E$ is derived as an approximation from the simulation data reported in [1, 6]. Figure 1 displays the CRP distribution within the Galaxy for energies of 1, $10^4$ , and $10^6$ TeV, illustrated in Figs. 1a, 1b, and 1c for a galactic slice at z = 0 kpc, and in Figs. 1d, 1e, and 1f for a slice at x = 0 kpc. The location of the Solar System within the Galaxy is indicated by a star at coordinates $(-8.2, 0, 0.2)$ kpc. Additionally, these graphs highlight the ratio of parallel to perpendicular components $D_{par} /D_{perp}$ of the diffusion tensor for the corresponding energies. CRP transport exhibits “preferred” propagation directions along the spiral arms of the galaxy, where $D_{par}$ significantly exceeds $D_{perp}$, often by one to two orders of magnitude. Conversely, in regions outside the spiral arms, the tensor components become practically equivalent, resulting in isotropic diffusion. The figures demonstrate that maximum elevated CRP number density occurs in regions where the $D_{par} /D_{perp}$ ratio is maximal. Meanwhile, the spatial distribution of CR sources contributes only minor modulation to the overall density distribution, as shown in Figs. 1g, 1h and 1i.\par
As energy increases beyond 1 PeV, the component
of the diffusion tensor responsible for perpendicular
transport grows, altering the transport dynamics outside the spiral arms from isotropic to anisotropic. This change facilitates the leakage of CRPs into the galaxy’s halo, particularly evident in Figs. 1e and 1f. We hypothesize that the mechanism of CRP leakage from the galactic disk into the halo is intricately linked to the anisotropic diffusion characteristics of CRPs and their complex energy dependent behavior. Establishing a correlation between our model and experimental data represents a challenging task, which we aim to address in future research.

\section{RESULTS. THE MODULATION
OF THE CR PROTON ENERGY SPECTRUM
IN THE “KNEE” REGION} \label{sec:Knee}

To evaluate the effectiveness of anisotropic trans-
port characteristics on the modulation of the proton
spectrum near 3 PeV, a simple power-law dependency $F = E^{-\gamma}$, where gamma is the spectral index, was
incorporated into the model as a source function; in
this instance, $\gamma = -2$. \par

We calculated the distribution of cosmic ray protons at the nodes of the grid, with the interval described earlier (see Section “Results. Calculation of CR’s Number Density”), throughout the volume of the galaxy. The characteristic dimensions were determined similarly to the parameters set in the model of the magnetic field [3]. Two scenarios were considered: a model of a flat halo with a minor semi-axis of 3 kpc and a major semi-axis of 17 kpc, and a scenario allowing for the free outflow of CR’s from the galactic disk, with a minor semi-axis of 10 kpc (“thick” halo).

\begin{figure*}
\includegraphics[width=0.95\textwidth]{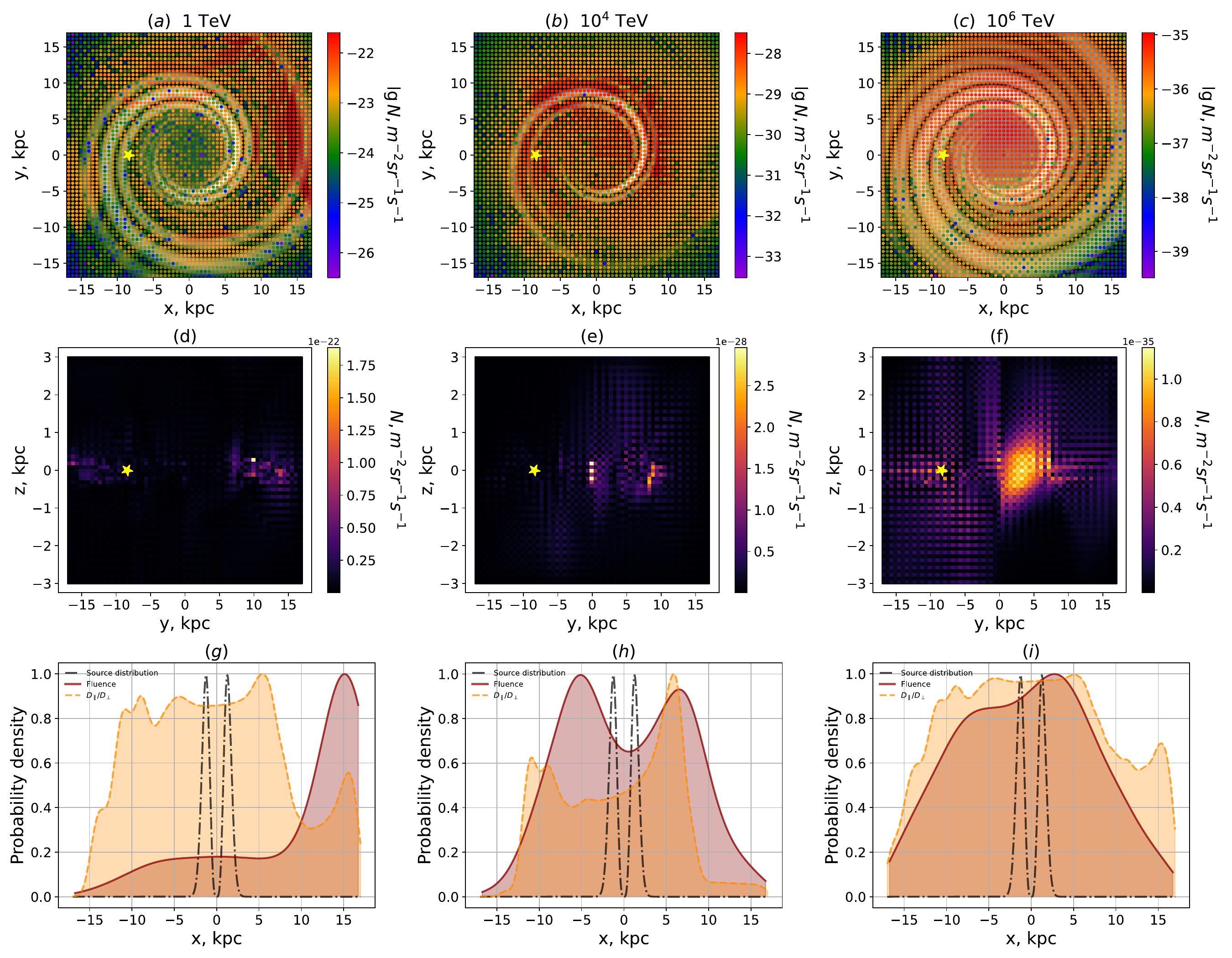}
\caption{\label{fig:dist}The CRP distribution within the Galaxy for energies of 1 TeV (a), (d), (g), $10^4$ TeV (b), (e), (h), and $10^6$ TeV (c), (f), (i) for
a galactic slice at z = 0 kpc (a), (b), (c) and for a slice at x = 0 kpc (d), (e), (f). Here the location of the Solar System is indicated
by a star. (g), (h), (i) Display normalized distribution densities for the CRPs, sources, and $D_{par} /D_{perp}$ , where the maximum value is taken as one.}
\end{figure*}

The implemented magnetic field model accounts for the exponential decay of the magnetic field’s absolute magnitude with distance from the galactic disk (for more details, see [3]). For each scenario, the energy spectrum was calculated ranging from 1 TeV to 1 EeV, with a step of one quarter-order of magnitude. For the spatial position of the Solar System within the galaxy, the constructed spectrum is presented in Fig. 2a. \par

The obtained spectrum was approximated using a
standard power function commonly employed to
describe a cutoff in the CR spectrum [8]:

\begin{eqnarray}
F_s(\gamma_1, \gamma_2, N_0, E_0, E) = N_0\left( \frac{E}{E_0} \right)^{-\frac{\gamma_1}{2}}&&\times\nonumber\\
\left( \frac{E}{E_0} \right)^{-\frac{\gamma_2}{2}} \left[ \frac{\left(E_0/E\right)^{\frac{s}{2}} + \left(E_0/E\right)^{\frac{s}{2}}}{2}\right]^{\frac{\gamma_1 - \gamma_2}{s}}
\label{eq:dist}
\end{eqnarray}

Here, $\gamma_1$ and $\gamma_2$ represent the spectral slope indices before and after the break, respectively, $E_0$ is the break position, $N_0$ is the number density at he break point, and $s$ is the smoothing parameter. To determine the optimal parameters of the employed function, a minimization method was applied using five free parameters $( \gamma_1, \gamma_2, N_0, E_0, E)$. For effective searching of the absolute minimum, the method of differential evolution was utilized. \par

For the position of the Solar System, the best parameters were found to be a break position $E_0$ at around 4 PeV, $s = 2.4$ , and $\Delta\gamma = 0.64$. The irregularities in the CR spectrum structure are due to the characteristics of the finite difference scheme calculations. This error is a systematic issue of the method used and is represented in the $\Delta\gamma$ uncertainties shown in Fig. 2b. These outliers decrease as the number of steps in the modeled volume increases and do not affect the spatial position of the break in the spectrum or the values of $\gamma_1$ and $\gamma_2$.

\begin{figure*}
\includegraphics[width=0.85\textwidth]{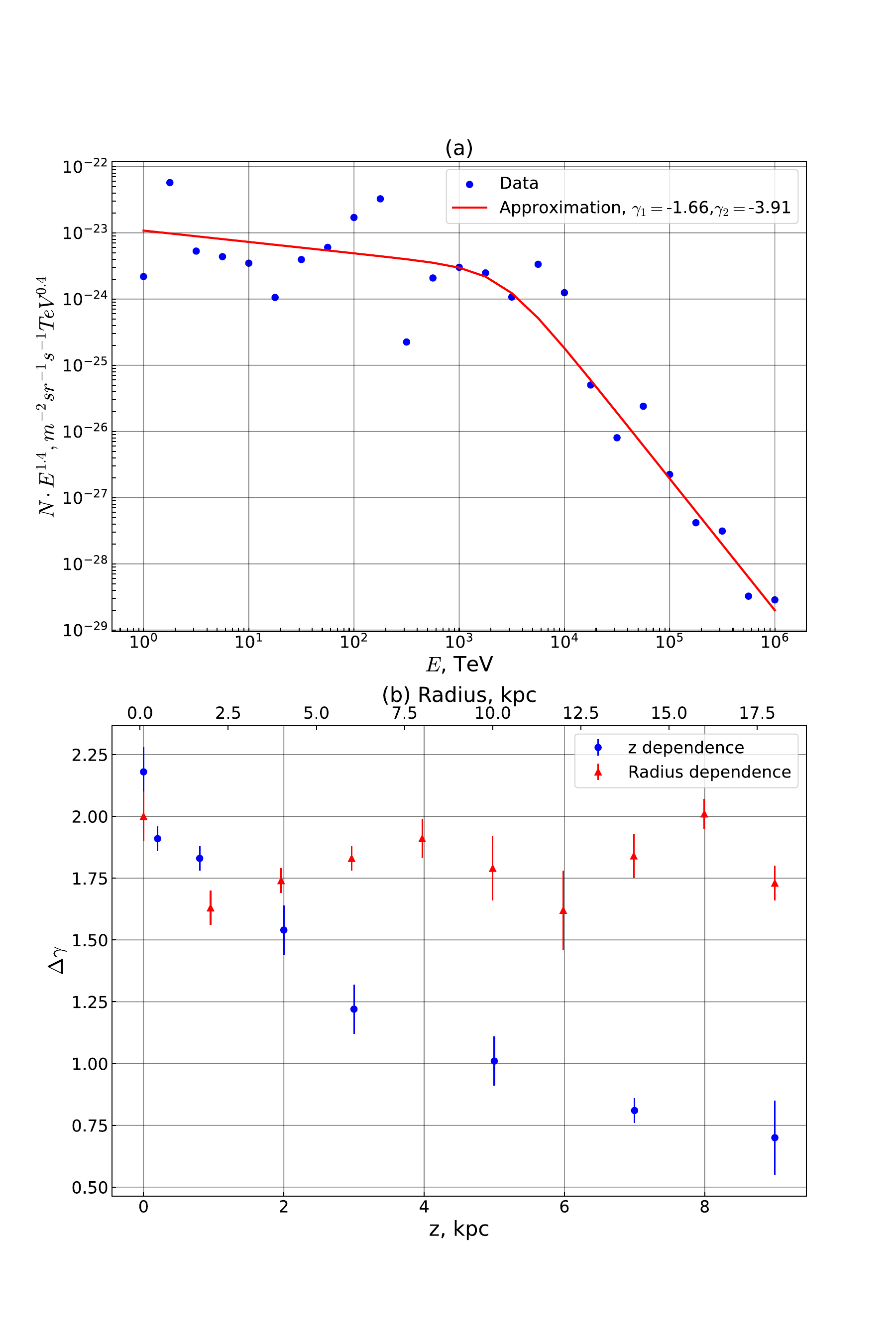}
\caption{\label{fig:dist}The dependence of the CRPs on energy (a), the spatial dependence of $\Delta\gamma$ on the radius R within the galaxy at a fixed $z = 0.2$ kpc, and z-dependence.}
\end{figure*}

A similar procedure was conducted for points in the galaxy different from the position of the Solar System, with two directions of interest being the dependence of $\Delta\gamma$ on the radius of the galaxy in the plane at $z = 0.2$ kpc and the z-dependence in a “thick” halo scenario $(x = y = 0 kpc)$. The results are presented in Fig. 2b. The highest $\Delta\gamma$ is observed in the arms of the galaxy, decreasing in the regions between them (where $D_{par} /D_{perp}$ decreases). Conversely, as one moves away from the disk of the galaxy, $\Delta\gamma$ decreases quite rapidly, and for $z > 6$, describing the spectrum with a simple power function without a break becomes acceptable, due to the increased uniformity of the field used in the model. The best function describing the modulation becomes a simple power function with $\gamma = –2$. Thus, it can be stated that anisotropic transport significantly influences the observed spectrum and its dependence on spatial position within the Galaxy. The results obtained from our model align well with the latest experimental data, particularly from the LHAASO collaboration [9], which indicate a spatial dependency of the cosmic ray knee.
\section{\label{sec:conclusion}CONCLUSIONS}

A mathematical model was developed to describe
the anisotropic transport of cosmic rays (CRs) with a
general tensor form, and a software package was made
to simulate the CR proton number density throughout
the galaxy, considering the realistic structure of the
magnetic field and the distribution of sources. This
model describes the modulation effect on the energy
spectrum of CR protons, with the break position
aligning well with experimental data. The characteristics of the break vary significantly depending on the
spatial position within the galaxy, which is consistent
with the latest experimental findings and underscores
the significant influence of CR transport in interpreting experimental data.

\section*{\label{sec:conclusion}CONFLICT OF INTEREST}
The authors of this work declare that they have no conflicts of interest.

\end{document}